\documentclass[
reprint,
superscriptaddress,
showpacs,
preprintnumbers,
nofootinbib,
nobibnotes,
amsmath,
amssymb, 
aps,
prd,
floatfix
]{revtex4-1}

\usepackage[utf8]{inputenc}
\usepackage[normalem]{ulem}
\usepackage{graphicx}
\usepackage{dcolumn}
\usepackage[colorlinks=true,allcolors=purple]{hyperref}
\usepackage{url}
\usepackage{enumerate}

\usepackage{slashed,multirow,relsize,soul,feynmp-auto,tikz}
\usepackage{color}
\usepackage{mathrsfs} 
\usepackage{amsmath}
\usepackage{cancel}
\usepackage{bbold}
 \usepackage{mathrsfs}
\usepackage{braket}
\usepackage{physics}
\usepackage{multirow}
\usepackage[capitalize]{cleveref}
\usepackage{xspace}

\usepackage{fontawesome} 
\definecolor{nicegreen}{rgb}{0., 0.75, 0.46}

\definecolor{MH}{rgb}{0.0,0.6,9}
\definecolor{palatinate}{rgb}{0.494, 0.192, 0.482}
\definecolor{blue-violet}{rgb}{0.33, 0.17, 0.89}

\renewcommand{\phi}{\varphi}

\begin{document}
\title{Decaying sterile neutrinos at short baselines}

\author{Matheus Hostert}
\email{mhostert@g.harvard.edu}
\affiliation{Department of Physics \& Laboratory for Particle Physics and Cosmology, Harvard University, Cambridge, MA 02138, USA}
\author{Kevin J. Kelly}
\email{kjkelly@tamu.edu}
\author{Tao Zhou}
\email{taozhou@tamu.edu}
\affiliation{Department of Physics and Astronomy, Mitchell Institute for Fundamental Physics and Astronomy, Texas A\&M University, College Station, TX 77843, USA}

\date{\today}

\begin{abstract}
Long-standing anomalous experimental results from short-baseline neutrino experiments have persisted for decades. These results, when interpreted with one or more light sterile neutrinos, are inconsistent with numerous null results experimentally. However, if the sterile neutrino decays en route to the detector, this can mimic $\nu_\mu \to \nu_e$ oscillation signals while avoiding many of these external constraints. We revisit this solution to the MiniBooNE and LSND puzzles in view of new data from the MicroBooNE experiment at Fermilab.
Using MicroBooNE's liquid-argon time-projection chamber search for an excess of $\nu_e$ in the Booster beam, we derive new limits in two models' parameter spaces of interest: where the sterile neutrino decays (I) via mixing with the active neutrinos, or (II) via higher-dimensional operators.
We also provide an updated, comprehensive fit to the MiniBooNE neutrino- and antineutrino-beam data, including appearance ($\nu_e$) and disappearance ($\nu_\mu$) channels.
Despite alleviating the tension with muon neutrino disappearance experiments, we find that the latest MicroBooNE analysis rules out the decaying sterile neutrino solution in a large portion of the parameter space at more than $99\%$~CL.
\end{abstract}

\maketitle

\section{Introduction}

The short-baseline neutrino anomalies present an unresolved puzzle in particle physics. 
In particular, the MiniBooNE low-energy excess stands at a significance of $4.8\sigma$~\cite{MiniBooNE:2018esg,MiniBooNE:2020pnu} with no compelling explanation within the Standard Model (SM).
At face value, these results are compatible with new oscillations driven by eV-scale sterile neutrinos in a cohesive picture with the observed LSND excess~\cite{LSND:1996ubh,LSND:2001aii}.
In practice, however, the global picture remains inconclusive due to significant tension between appearance ($\nu_\mu \to \nu_e$) and disappearance ($\nu_e \to \nu_e$ and $\nu_\mu \to \nu_\mu$) experiments~\cite{Dentler:2018sju,Diaz:2019fwt}.
Radioactive source experiments show evidence for $\nu_e$ disappearance at the level of $20\%$, an anomaly commonly referred to as the Gallium anomaly~\cite{Barinov:2022wfh} (see Refs.~\cite{Giunti:2022btk,Brdar:2023cms}).
However, such a large effect is in direct contradiction with reactor~\cite{DANSS:2018fnn,STEREO:2019ztb,PROSPECT:2020sxr} and solar neutrino~\cite{Goldhagen:2021kxe} experiments. 
Searches for $\nu_\mu$ disappearance have also been used to set strong limits on oscillation models. 
MINOS/MINOS+, for instance, is responsible for a significant\footnote{The limits, however, appear to be too strong and have been debated in the literature~\cite{Louis:2018yeg,Diaz:2019fwt}.} amount of the internal tension~\cite{MINOS:2017cae}.
IceCube has recently revealed a mild preference for resonant eV-sterile neutrino $\nu_\mu$ disappearance in matter at the $95\%$ CL~\cite{MEOWS2024}.
While neutrino data ultimately disagree on the need for new physics, it is clear that minimal eV-scale sterile neutrino models are unlikely to be the final answer.
This is also strongly corroborated by cosmology~\cite{Planck:2018vyg,Hagstotz:2020ukm,Adams:2020nue}, which rules out the new oscillation picture at a high significance.
In view of that, we turn to new potential resolutions\footnote{For a more thorough discussion of these hints and tensions, as well as other proposed solutions, we refer the interested reader to Ref.~\cite{Acero:2022wqg}.} to the short-baseline anomalies and, in particular, to models where sterile neutrinos induce flavor transitions not only through oscillation but also through their decays.

Light sterile neutrinos that decay to (visible) electron-neutrinos were first proposed as an explanation to the LSND excess~\cite{Palomares-Ruiz:2005zbh} and were later revisited after the MiniBooNE results~\cite{Bai:2015ztj,deGouvea:2019qre,Dentler:2019dhz}.
The main difference with respect to the oscillation-only scenario is the fact that the $\nu_\mu \to \nu_e$ appearance signal is generated by decays, and therefore, it is only suppressed by the square as opposed to the fourth power of small mixing-matrix elements.
This dependence accommodates the anomalies with much smaller values of $|U_{\mu 4}|$, the mixing element between the mostly-sterile neutrino and the muon flavor, and effectively evades limits from $\nu_\mu$ disappearance searches.
This is in contrast with invisible neutrino decays~\cite{Moss:2017pur}, where the LSND and MiniBooNE results are still explained by oscillations.
The new force behind neutrino decay can also help reconcile light sterile neutrinos with cosmology through the so-called secret-interaction mechanism~\cite{Dasgupta:2013zpn,Hannestad:2013ana,Farzan:2019yvo,Cline:2019seo}.
This resolution is based on the fact that neutrino mixing can be strongly suppressed in the early universe due to strong self-interactions between sterile neutrinos.
The same force that induced neutrino decay could be responsible for such effects.

In this article, we revisit the MiniBooNE excess in view of the MicroBooNE data in a few variations of decaying sterile neutrino models. 
We perform a full fit to the MiniBooNE $\nu_e$ and $\nu_\mu$ samples in neutrino and antineutrino mode, building on previous literature by using the $\nu_e$ and $\nu_\mu$ Monte-Carlo provided by the collaboration and accounting for the kinematics of neutrino decay, the effect of neutrino disappearance, and of neutrino regeneration.
For completeness, we also present fits to these MiniBooNE data in the standard stable sterile neutrino scenario.
We then derive the MicroBooNE constraints on the decaying-sterile neutrino hypothesis by adapting MicroBooNE's $\nu_e$ template search~\cite{MicroBooNE:2021tya,MicroBooNE:2021nxr,MicroBooNE:2021pvo,MicroBooNE:2021wad}, extending the work in Ref.~\cite{Arguelles:2021meu}.
These complement the existing studies of sterile neutrinos at MicroBooNE performed by phenomenologists~\cite{Arguelles:2021meu,Denton:2021czb} as well as by the MicroBooNE collaboration~\cite{MicroBooNE:2022sdp}.
We focus on decaying scenarios without $\nu \to \overline{\nu}$ transitions, evading strong constraints from solar antineutrino searches~\cite{Hostert:2020oui}.

This paper is organized as follows.
In \cref{sec:model}, we present the decaying-sterile neutrino models and discuss the signal rate for effective flavor transitions at short-baseline experiments.
We then describe our MiniBooNE and MicroBooNE fits in \cref{sec:fits}, present our results in \cref{sec:discussion}, and conclude in \cref{sec:conclusions}.

\section{Decaying Steriles}
\label{sec:model}

In this section, we present the decaying-sterile neutrino models. 
We analyze sterile neutrino decays to scalar particles $\phi$, although phenomenologically, decays to vector particles are analogous.
We consider two cases: (I) the scalar particle couples exclusively to the singlet $\nu_s$ flavor, where the decay is induced exclusively by flavor mixing (analogous to the model in Ref.~\cite{Dentler:2019dhz}), and (II) the scalar couples to neutrinos exclusively via a dimension-five operator involving the electron neutrino flavor (analogous to the models in Refs.~\cite{Palomares-Ruiz:2005zbh,deGouvea:2019qre}).
In scenario (II), the mass eigenstate $\nu_4$ mixes only with the muon flavor but decays to $\nu_e$ states due to the higher-dimensional operator.
We also assume neutrinos to be Dirac and $\phi$ to be stable, avoiding $\nu \to \overline{\nu}$ transitions at short baselines and from the Sun. 

\begin{figure}[t]
    \centering
    \includegraphics[width=0.33\textwidth]{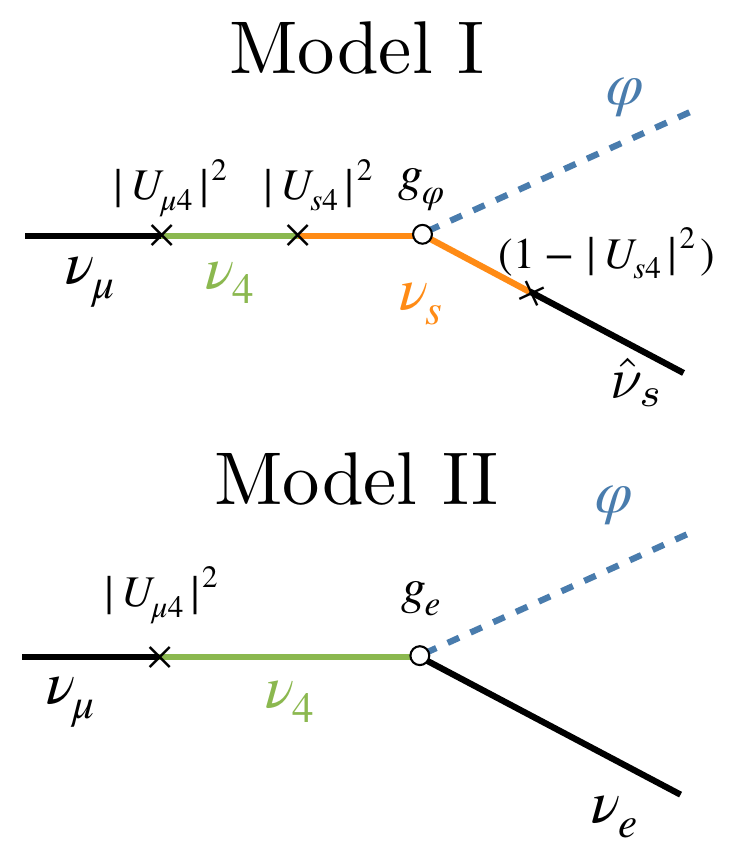}
    \caption{A depiction of the effective $\nu_\mu \to \nu_e$ transition induced by the two sterile neutrino decay models considered in this work.
    Top: a sterile neutrino mixed with electron and muon flavors -- \cref{eq:modelI}.
    Bottom: a sterile neutrino mixed only with muon flavor that decays via a higher-dimensional operator -- \cref{eq:modelII}.\label{fig:diagram}}
\end{figure}

\begin{figure*}[t]
     \centering
     \includegraphics[width=0.49\textwidth]{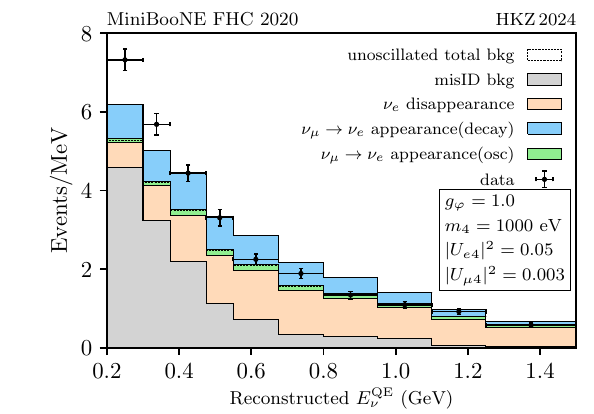}
     \includegraphics[width=0.49\textwidth]{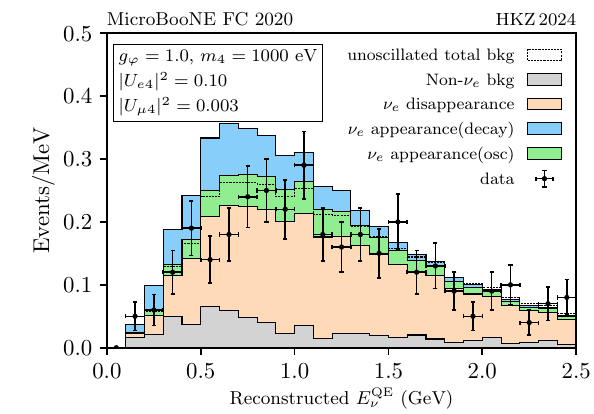}
     \caption{Excess of electron-like events at MiniBooNE(left) and MicroBooNE(right) in neutrino mode. The intrinsic $\nu_e$ contribution to the background is shown as a salmon color and differs from the baseline MiniBooNE expectation (dashed black) due to $\nu_e$ disappearance. The green-filled histograms represent the predicted $\nu_e$ appearance from oscillation, while the blue ones show the contribution from decay (Model-I).
     \label{fig:spectra}}
 \end{figure*}

\subsection{Model-I: mixing with all flavors}

We introduce a Dirac sterile neutrino $\nu_s$ and a light scalar $\phi$.
The Lagrangian is given by
\begin{equation}\label{eq:modelI}
    - \mathscr{L} \supset g_\phi \overline{\nu_s} \nu_s \phi + \sum_{\alpha, \beta} m_{\alpha \beta} \overline{\nu}_\alpha\nu_\beta,
\end{equation}
where $g_\phi$ is the coupling constant and $m_{\alpha \beta}$ the mass matrix for neutral leptons.
Upon diagonalization of $m_{\alpha \beta}$, the physical neutrino mass states $\nu_i = U_{\alpha i}^* \nu_\alpha$ interact with the scalar particle via
$g_\phi U_{si}^* U_{sj} \overline{\nu_i} \nu_j + \text{ h.c.}$
In this model, the fourth mass eigenstate can lead to two new effects at accelerator neutrino experiments.
The first is the usual oscillation induced by the mass splitting $\Delta m_{41}^2 = m_4^2 - m_1^2 \simeq m_4^2$.
This leads to $\nu_\mu \to \nu_e$ appearance at the level of $|U_{e4}|^2 |U_{\mu 4}|^2$ in much the same way as the minimal eV-sterile neutrino models albeit with an imaginary phase induced by the finite lifetime of the neutrinos.
The second comes from the decay of the fourth mass eigenstate, $\nu_4 \to \hat{\nu}_s \phi$, where $\hat{\nu}_s$ is a linear combination of the three lighter mass eigenstates.
This can also lead to an apparent $\nu_\mu \to \nu_e$ appearance, now at the level of $|U_{\mu 4}|^2$, provided $\hat{\nu}_s$ has a large overlap with the $\nu_e$ flavor state. The top diagram of~\cref{fig:diagram} demonstrates this process.

Neutrinos (antineutrinos) in the accelerator experiments are, to a good approximation, left-handed (right-handed) chiral states, i.e., eigenstates of the Weak interaction. 
The same is true for the decaying steriles, which are mostly of left-handed (right-handed) type $\nu_4^-$ ($\nu_4^+$) in neutrino (antineutrino) modes.
Its decays may be helicity-conserving ($\nu_4^- \to \nu^- \phi$) or helicity-flipping ($\nu_4^- \to \nu^+ \phi)$. 
In the latter case, the daughter neutrinos are invisible as they do not interact through the Weak force.
Therefore, we will only be interested in the helicity-conserving decays.
We further assume that this is the only type of decay allowed in the model.\footnote{This is easily achieved in parity-violating scenarios. For instance, if only left-handed mass eigenstates mix with the sterile flavor, then $\nu_{1,2,3}^+$ cannot be produced in the decays of $\nu_4^-$.}
As it turns out, these decays tend to produce higher energy daughter neutrinos than their helicity-flipping counterparts.
We return to this point in \cref{sec:event_rates,app:HF}.

The width of the fourth mass eigenstate in the laboratory frame is given by
\begin{equation}
    \Gamma^{\rm (I)}_{\nu_4}  = \Gamma_{\nu_4 \to \hat{\nu}_s \phi} = |U_{s4}|^2(1 - |U_{s4}|^2)\frac{\alpha_\phi}{4} \frac{m_4^2}{E_4},
\end{equation}
where $\alpha_\phi = g_\phi^2/4\pi$.
We neglected the mass of the daughter neutrinos and defined the low-energy flavor state 
\begin{equation}
    \ket{\hat{\nu}_s} = \frac{1}{\left(\sum_{k=1}^3 |U_{s k}|^2\right)^{1/2} } \sum_{i=1}^3 U_{si}^* \ket{\nu_i},
\end{equation} 
which is normalized according to the unitary mixing matrix $U$.

At the expense of fine-tuning, we assume the scalar particle to be lighter than the three lightest neutrinos. 
In other words, $m_\phi \leq m_{\rm lightest}$ with $m_{\nu_{\rm lightest}}$ the mass of the lightest neutrino eigenstate. 
This choice forbids $\phi \to \nu \overline{\nu}$ decays.
We also note that limits from lepton-number-violating processes like neutrino-less double beta decay do not apply as lepton number is conserved. 
Meson decays like $\pi^+, K^+ \to \ell^+ \nu_s \phi$ are suppressed as the scalar interaction conserves flavor and the mixing matrix $U$ is unitary.

\subsection{Model-II: No electron flavor mixing}

Now we turn to the effective model in Refs.~\cite{Palomares-Ruiz:2005zbh,deGouvea:2019qre}.
At low energies, the particle content is identical to the one above, but the scalar interactions with $\nu_s$ are replaced by the following higher-dimensional operator,
\begin{equation}
\label{eq:modelII}
    -\mathscr{L} \supset \frac{C_e}{\Lambda} \overline{L_e} \tilde{H} \phi \nu_4 + \text{ h.c.} \xrightarrow{\slashed{\rm EW}}
 g_e \overline{\nu_e} \phi \nu_4 + \text{ h.c.} ,
\end{equation}
where we have defined $g_e = C_e v/\sqrt{2}\Lambda$ as the low-energy coupling, $L_e$ is the SU$(2)_L$ electron doublet, and $\tilde{H}$ is the charge-conjugate Higgs field.
Because the operator above leads to $\nu_4 \to \nu_e \phi$ decays, it replaces the need for a non-zero mixing between the fourth mass eigenstate and the electron flavor.
Therefore, in the current scenario, it is possible to set $|U_{e4}| = 0$ and to decouple the decaying sterile explanation of the $\nu_\mu \to \nu_e$ anomalies from $\nu_e \to \nu_e$ disappearance experiments.
In addition, $|U_{e4}| = 0$ also shuts off $\nu_\mu \to \nu_e$ oscillations, implying that the MiniBooNE signal is exclusively due to the decays of $\nu_4$.

Neglecting contributions from the mixing, the sterile decay width is given by
\begin{align}
    \Gamma^{\rm (II)}_{\nu_4}  &= \Gamma_{\nu_4 \to \nu_e \phi} = \frac{\alpha_e}{4} \frac{m_4^2}{E_4},
\end{align}
where $\alpha_e = g_e^2/4\pi$.
We will be particularly interested in the value of the range $g_e \sim 10^{-3}$, implying a new physics scale of about $\Lambda/C_e \sim 10^2$~TeV.
In this model, when $|U_{e4}| = |U_{\tau 4}| = 0$, the low-energy states $\hat{\nu}_e$ and $\hat{\nu}_\tau$ are, in fact, identical to the full flavor states $\nu_e$ and $\nu_\tau$.
Similarly to before, we assume that only helicity-conserving decays are allowed, thereby fixing the chiral structure of the scalar interaction.

The interaction in \cref{eq:modelII} violates flavor and is consequently constrained by rare processes at low energies.
In particular, the most direct limits are given by
\begin{align}
    \pi^+,K^+ &\to \nu e^+\phi \quad \to\quad  &g_e^2 &< 1.9 \times 10^{-6},
    \\
    \pi^+,K^+ &\to \nu_e \mu^+\phi \quad \to \quad &g_e^2 |U_{\mu 4}|^2 &< 1.9 \times 10^{-7}
\end{align}
at the $90\%$ CL~\cite{Pasquini:2015fjv}.
The mixing $|U_{e4}|^2$ is also constrained by kink searches in beta decay~\cite{Shrock:1980vy,Dragoun:2015oja,Abdurashitov:2017kka,KATRIN:2022ith,KATRIN:2022spi}.
As before, other stringent limits from cosmology will apply, constraining both the mixings $|U_{e 4}|, |U_{\mu 4}|$ and the scalar-neutrino coupling $g_e$.
In this model, the secret-interaction mechanism is not immediately applicable.
Evading thermalization of $\nu_4$ or ensuring that neutrinos free stream at late times will require further model building on the particle physics and/or cosmology side.

\begin{figure*}[t]
    \centering
    \includegraphics[width=0.49\textwidth]{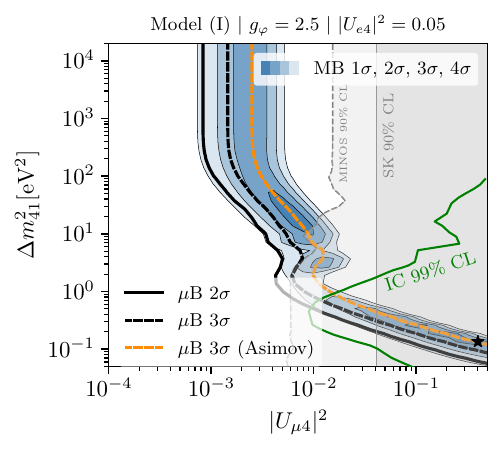}
    \includegraphics[width=0.49\textwidth]{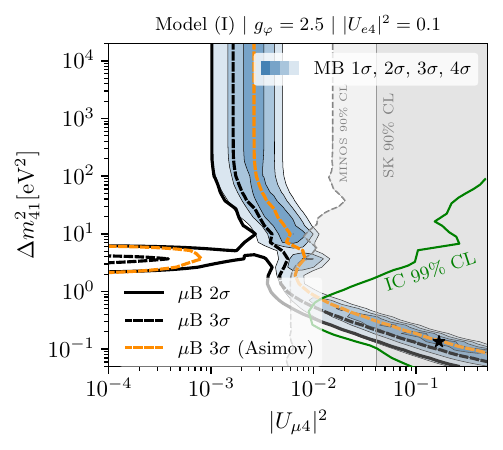}
    \caption{The parameter space of Model (I) for specific choices of $|U_{e4}|^2$ and $g_\phi$.
    On the left, we set $|U_{e4}|^2 = 0.05$, corresponding approximately to the $90\%$ C.L constraint on this parameter from solar neutrino experiments, and $g_\phi = 2.5$. On the right, we set $|U_{e4}|^2 = 0.1$, corresponding approximately to the best fit of the BEST experiment, and the same $g_\phi$.
    The MicroBooNE constraints are shown as black solid(dashed) lines at $2\sigma$($3\sigma$).
    The Asimov sensitivity expectation of MicroBooNE at $3\sigma$ CL is shown for comparison as a dashed orange line.
    The green line shows the $99\%$ CL IceCube exclusion region~\cite{weigel-2024}, neglecting $\nu_\mu$ regeneration effects. The grey and light-grey regions show the $90\%$ CL excluded by Super-Kamikande~\cite{Super-Kamiokande:2014ndf} and MINOS/MINOS+~\cite{MINOS:2017cae}.
    }
    \label{fig:modelIfit_fixedUe4}
\end{figure*}

\subsection{Short-Baseline Event Rates} \label{sec:event_rates}

We now discuss our treatment of $\nu_\alpha \to \nu_\beta$ flavor transitions.
We start with the flux of $\nu_\beta$ neutrinos at the detector location as a function of the true neutrino energy, $\Phi_\beta(L, E_\nu)$.
Neglecting the mass of $\nu_4$ in the kinematics,
\begin{equation}\label{eq:rate_relation}
    \Phi_{\beta}(L, E_\nu) = \sum_\alpha\int_{E_\nu}^\infty \dd E_4   P_{\alpha\beta}(L,E_4,E_\nu) \Phi_\alpha (L, E_4),
\end{equation}
where $P_{\alpha\beta}(L,E_4,E_\nu)$ is an effective flavor transition probability and $\Phi_\alpha(L, E_4)$ is the flux of $\nu_\alpha$ neutrinos evaluated at the $\nu_4$ energy.
Since the flavor evolution of decay products does not interfere with that of the parent neutrinos, we can treat the two components separately,
\begin{align}\label{eq:fullprob}
    P_{\alpha\beta}(L, E_4, E_\nu) &= P_{\alpha \beta}^{\rm dec}(L, E_4, E_\nu) S^{\rm dec}_{\alpha \beta}(E_4, E_\nu) 
    \\
    \nonumber & \quad + P_{\alpha\beta}^{\rm osc}(L, E_\nu) \delta(E_4 - E_\nu).
\end{align}
For oscillations, the delta function ensures energy conservation, while for decays, the parent and daughter energy will differ.
The energy-migration factor $S^{\rm dec}_{\alpha \beta}(E_4, E_\nu)$ accounts for the effect of energy degradation on the detection rate.

By explicitly time-evolving the neutrino states with real and imaginary phases and neglecting phase differences between light mass eigenstates, we find
\begin{widetext}
\begin{equation}\label{eq:Posc}
    P_{\alpha\beta}^{\rm osc} (L, E_\nu) = \delta_{\alpha \beta} - 2 \delta_{\alpha \beta} |U_{\alpha 4}U_{\beta4}| \left[ 1 - e^{-\frac{L}{2 L_{\rm dec}}} \cos\left(\pi\frac{L}{L_{\rm osc}}\right)\right]
    + |U_{\alpha 4}U_{\beta4}|^2 \left[ 1 - 2 e^{-\frac{L}{2 L_{\rm dec}}} \cos\left(\pi\frac{L}{L_{\rm osc}}\right)  + e^{-\frac{L}{L_{\rm dec}}}\right],
\end{equation}
\end{widetext}
where $L_{\rm osc} = 2\pi E_\nu/\Delta m^2_{41}$ and $L_{\rm dec} = 1 / \Gamma_{\nu_4}$.
Note that for $L=0$, no flavor conversion is possible, $P_{\alpha\beta} (L=0, E_\nu) = \delta_{\alpha\beta}$.
In the fast oscillation regime, the probability simplifies to $P_{\alpha\beta}^{\rm osc}(L \gg L_{\rm osc}, E_\nu)  = (\delta_{\alpha \beta} - |U_{\alpha 4} U_{\beta 4}|)^2 + |U_{\alpha 4} U_{\beta 4}|^2 e^{-\frac{L}{L_{\rm dec}}}$.
Physically, the two terms correspond to the decoherence of $\ket{\nu_\alpha}$ state into a low-energy $\ket{\hat{\nu}_\alpha} = \sum_{i = 1,2,3} U_{\alpha i}^* \ket{\nu_i}$ flavor state and a massive $\nu_4$ eigenstate.
The former is independent of the properties of $\nu_4$ while the latter is appropriately suppressed by the probability of $\nu_4$ to decay between the source and the detector.

\begin{figure}[t]
    \centering
    \includegraphics[width=0.49\textwidth]{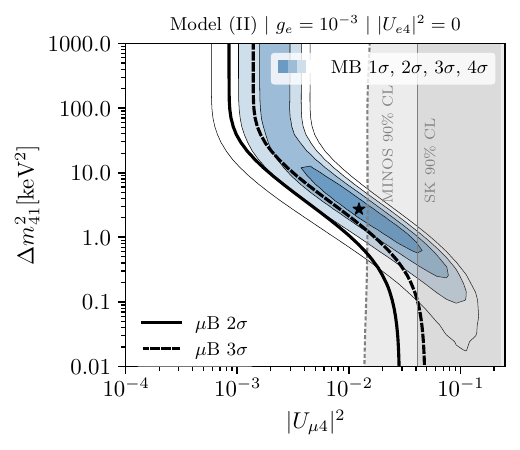}
    \caption{The parameter space of Model (II) for $g_e = 10^{-3}$ with $|U_{e4}|^2 = 0$.
    The production of $\nu_4$ is set by $|U_{\mu 4}|^2$, which can be small enough to evade $\nu_\mu$ disappearance constraints.
    The $2\sigma$ ($95.4\%$ CL) and $3\sigma$ ($99.7\%$ CL) MicroBooNE constraints are shown in black solid and dashed lines, respectively.
    \label{fig:modelIIfit_fixedUe4}
}
\end{figure}

We now turn to the decay term that accounts for the visible decay products of $\nu_4$ decays.
For a fixed $\nu_4$ energy, the probability of detecting a neutrino of flavor $\nu_\beta$ and of energy $E_\nu$ in the detector is
\begin{align}\label{eq:Pdec}
    P_{\alpha\beta}^{\rm dec}(L, E_4, E_\nu) &= |U_{\alpha 4}|^2\frac{|\bra{\hat{\nu}_{s}}\ket{\nu_\beta}|^2}{|\bra{\hat{\nu}_{s}}\ket{\hat{\nu}_{s}}|^2}  (1 - e^{-\frac{L}{L_{\rm dec}}}),
\end{align}
where the flavor projection is given by $\bra{\hat{\nu}_{s}}\ket{\nu_\alpha} = \frac{|U_{s4}|^2|U_{\alpha 4}|^4}{1-|U_{s4}|^2}$.
The energy migration of neutrinos is accounted for by
\begin{align}
    S^{\rm dec}_{\alpha \beta}(E_4, E_\nu) &=
    \left(\frac{1}{\Gamma_{\nu_4}} \frac{\dd \Gamma_{\nu_4 \to \nu \phi}}{\dd E_\nu} \right),
\end{align}
where the last factor above depends on the dynamics of the decay.
For models I and II, for helicity-conserving decays\footnote{For completeness, we briefly explore the helicity-flipping case -- where $\dd\Gamma/\dd E_\nu \propto 1 - E_\nu/E_4$ -- in~\cref{app:HF}.}, we have
\begin{equation}\label{eq:dPdX}
    \frac{1}{\Gamma_{\nu_4}} \frac{\dd \Gamma_{\nu_4 \to \nu \phi}}{\dd E_\nu} = \frac{ 2E_\nu}{E_4^2}.
\end{equation}
Note that we implicitly assumed that the geometric acceptance of the detector to the daughter neutrinos is $100\%$, which holds to a very good approximation for light $\nu_4$.
The typical opening angle between a daughter particle and its parent $P$ in a two-body decay is $\theta_\nu {\sim}1/\gamma = m_P/E_P$, which is kept far below $\sim (300 \text{ keV})/(300\text{ MeV}) = 10^{-3}$ in our analysis, to be compared with the detector solid-angle $\theta_{\rm MB}\sim 10$ m/$500$ m $= 2\times10^{-2}$.

In practice, \cref{eq:rate_relation} can be discretized as a migration matrix: for every $E_\nu$, the rate depends on a probability distribution of $E_4$ values, rendering the integral in \cref{eq:rate_relation} to a sum over $E_4$ bins.
This is the procedure we adopt in our analysis. 
To obtain the spectrum in reconstructed energy, then, one need only convolve the true event rate spectrum $R(E_\nu)$ with the experiment's smearing matrix $M(E_\nu^{\rm reco},E_\nu)$, that is, $R(E_\nu^{\rm reco}) = \int \dd E_\nu M(E_\nu^{\rm reco},E_\nu)R(E_\nu)$.

\renewcommand{\arraystretch}{1.3}
\begin{table*}[t]
    \centering
    \begin{tabular}{|l|c|c|c|c|c|c|}
     \hline
      New Physics Model & $|U_{e4}|^2$ & $|U_{\mu 4}|^2$ & $g_\phi$ or $g_e$ & $\Delta m_{41}^2$ & $\chi^2_{\rm MB} - \chi^2_{{\rm MB},{\rm Null}}$ & $\chi^2_{\mu{\rm B}} - \chi^2_{\mu{\rm B},{\rm Null}}$ 
     
     \\ \hline \hline
    
    Oscillations App only & \multicolumn{2}{|c|}{$\sin^22\theta_{e\mu} = 0.42$} & --- & $6.6 \times 10^{-2}$~eV$^2$ & $-32$ & $21$
    \\\hline
    Oscillations Full & $3.1\times10^{-2}$ & $4.2\times10^{-1}$ & --- & $0.18$ eV$^2$ & $-33$ & $24$
    \\ \hline\hline
    \multirow{3}{*}{Decay model (I)} & $2.0\times 10^{-2}$ & $1.1\times 10^{-2}$ & $3.0$ & $19$~eV$^2$ & $-33$ & $24$
    \\\cline{2-7}
    & $2.1\times 10^{-2}$ & $9.8\times 10^{-3}$ & $2.5$ (fixed) & $29$~eV$^2$ & $-33$ & $22$
    \\\cline{2-7}
     & $1.3\times 10^{-2}$ & $1.2\times 10^{-2}$ & $1.0$ (fixed) & $2.4\times 10^{2}$~eV$^2$ & $-33$ & $23$
    \\ \hline
    \multirow{2}{*}{Decay model (II)} & $0$ (fixed) & $1.2\times 10^{-2}$ & $1\times 10^{-3}$ (fixed) & $2.7$ keV$^2$ & $-35$ & $20$
    \\\cline{2-7}
    & $0.1$ (fixed) & $4.5\times 10^{-3}$ & $1\times 10^{-3}$ (fixed) & $8.9$ keV$^2$ & $-30$ & $21$
    \\
    \hline
    \end{tabular}
    \caption{The model parameters of our MiniBooNE ($\nu$ + $\overline{\nu}$) best-fit points in $3+1$ oscillation and decaying sterile neutrino models (I) and (II), imposing the constraint $|U_{\mu 4}|^2 < 0.1$. 
    In Model-I, the MicroBooNE best fit is given by $g_\phi = 3.0$, $\Delta m^2_{41} = 1.1$~eV$^2$, $|U_{e4}|^2 = 0.086$, and $|U_{\mu 4}|^2 = 1.3\times 10^{-4}$, with $\chi^2_{\mu{\rm B}} - \chi^2_{\mu{\rm B},{\rm Null}} = -2.8$.
    The null hypotheses have $\chi^2_{\rm MB} = 69$ and $\chi^2_{\rm \mu B} = 93$.
    \label{tab:bestfits}}
\end{table*}

\begin{figure*}
    \centering
    \includegraphics[width=\textwidth]{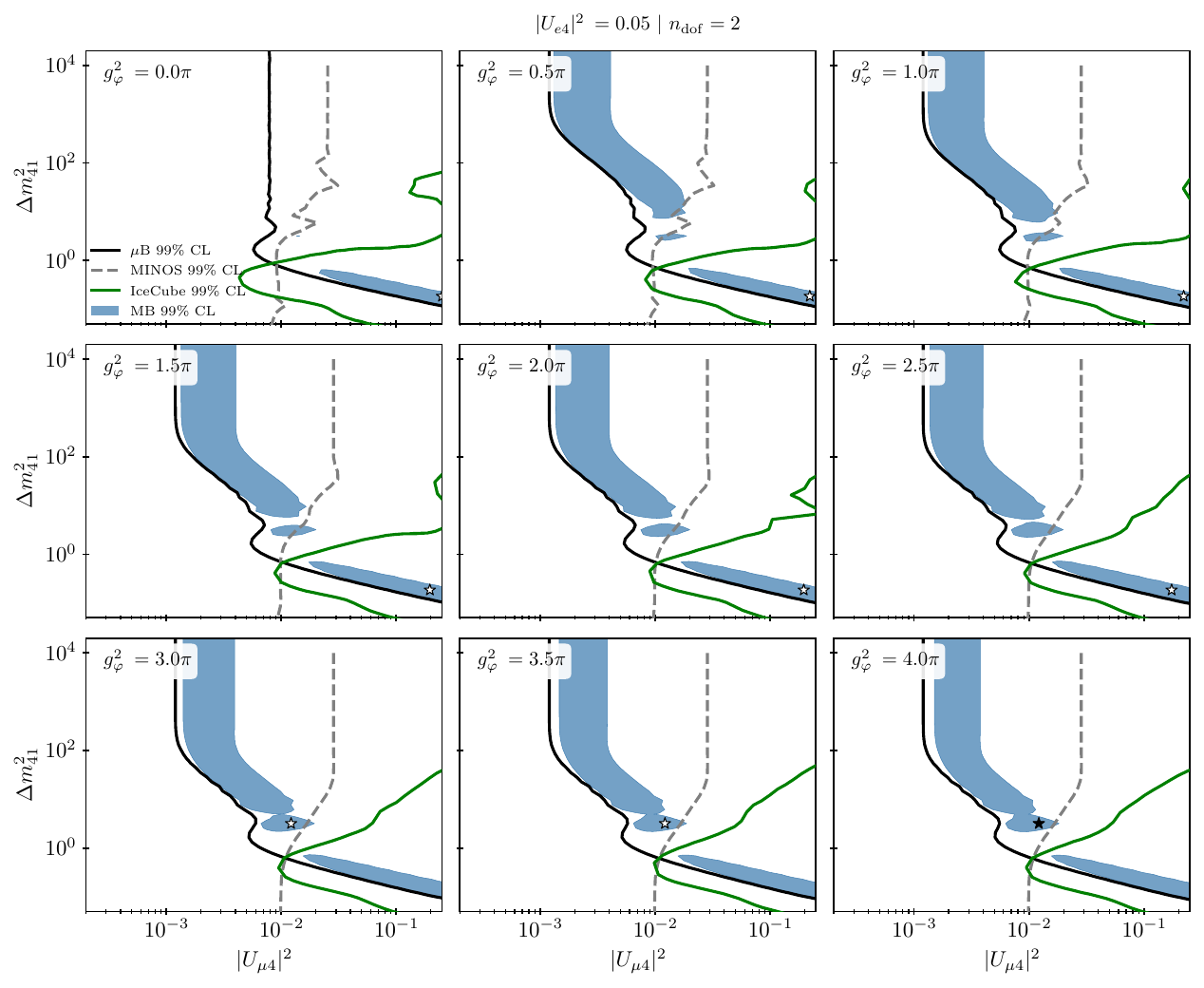}
    \caption{Comparison of $99\%$~CL exclusion limits set by MicroBooNE (black), MINOS (dashed grey), IceCube (green)~\cite{weigel-2024}, and $99\%$~CL preferred region by MiniBooNE (blue). We take 9 slices of three-dimensional parameter space of Model (I), fixing $|U_{e4}^2|=0.05$. Each panel corresponds to a fixed value of $g_\phi$ following the IceCube analysis in~\cite{weigel-2024}.
    Open and closed stars correspond to the local and global best-fit of points of MiniBooNE.
    \label{fig:IceCube_comparison}}
\end{figure*}

\begin{figure*}
    \centering
    \includegraphics[width=\textwidth]{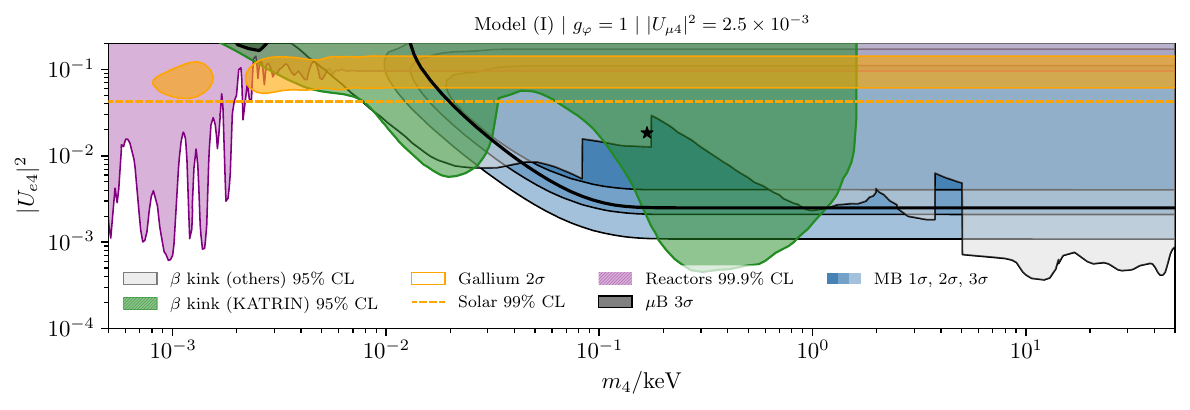}
    \includegraphics[width=\textwidth]{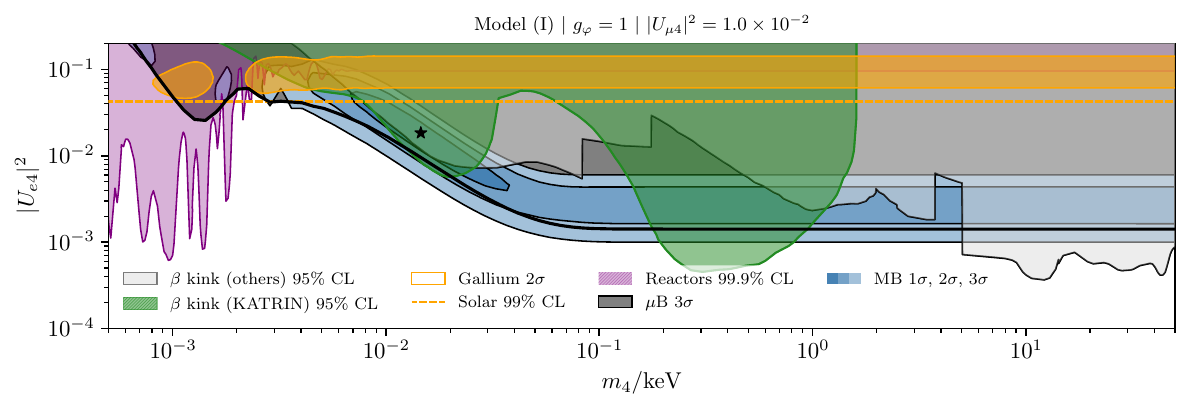}
    \caption{The parameter space of Model (I) for $g_\phi = 1$ fixing $|U_{\mu 4}|^2=2.5\times10^{-3}$ (top panel) and $|U_{\mu 4}|^2 = 10^{-2}$ (bottom panel).
    To obtain the MiniBooNE and MicroBooNE contours, we profile over $|U_{\mu 4}|^2$ constraining $|U_{\mu 4}|^2 < 0.04$ to respect limits from Super-Kamiokande~\cite{Super-Kamiokande:2014ndf}.
    Current MicroBooNE data excludes all of the $1\sigma$ and part of the $2\sigma$ regions of preference for MiniBooNE at the 3$\sigma$ level.
    Model (II) does not require a non-zero value for $|U_{e4}|^2$ to explain the MiniBooNE excess.
    In orange, we show the region preferred by the Gallium results~\cite{Barinov:2022wfh}, in purple, the region excluded at $99.9\%$ CL by reactor experiments~\cite{Declais:1994su,DANSS:2018fnn,DayaBay:2018yms,DoubleChooz:2019qbj,NEOS:2016wee,RENO:2018dro,Berryman:2020agd,Berryman:2021yan}, in green the $95\%$ CL exclusion set by the KATRIN experiment~\cite{KATRIN:2020dpx,KATRIN:2022ith,KATRIN:2022spi}, and in grey from other kink searches~\cite{Dragoun:2015oja,Abdurashitov:2017kka}.
    \label{fig:Ue4sq_vs_m4}
}
\end{figure*}

\section{ MiniBooNE and MicroBooNE fits}
\label{sec:fits}

Equipped with the flavor conversion rate calculation above, we performed a comprehensive fit to the MiniBooNE low-energy excess in the parameter space of Model-I and Model-II.
The fit was performed using the publicly-available MiniBooNE $\nu_\mu \to \nu_e$ Monte-Carlo events and covariance matrices.
We follow the fitting procedure adopted by the collaboration, including both appearance and disappearance effects~\cite{MiniBooNE:2022emn}.
We show the combined fit of neutrino-enhanced and antineutrino-enhanced modes from the 2020 MiniBooNE results~\cite{MiniBooNE:2020pnu}, fitting the $\nu_e$CC and $\nu_\mu$CC rate simultaneously.
The $\nu_e$CC rate contains $\nu_\mu \to \nu_e$ oscillation and decay events (neutrinos associated with the $\nu_\mu$ flux) as well as an intrinsic $\nu_e$ that can undergo $\nu_e \to \nu_e$ disappearance.
For $\nu_\mu$CC, we neglect eventual $\nu_e \to \nu_\mu$ oscillations as the intrinsic $\nu_e$ flux is far smaller than the $\nu_\mu$ one.
We also neglect wrong-sign neutrinos in a given mode, which is expected to be subdominant, especially for the neutrino-enhanced mode, which has the most statistically significant excess.

The energy degradation procedure for the MiniBooNE appearance channel goes as follows. For each parent neutrino from the MC record, we create a subset of possible daughter neutrinos with energy distribution according to~\cref{eq:dPdX}. Next, we calculate the flavor transition probabilities according to each daughter neutrino's baseline and energy, accounting for the shift in the total neutrino-nucleus cross-section and energy-dependent efficiency. We then bin the extended Monte-Carlo events in true daughter neutrino energy and migrate this true energy distribution to MiniBooNE's official reconstructed energy bins using the migration matrix extracted from the Monte-Carlo events. In the case of disappearance channels, we follow a similar approach.
For $\nu_e$ disappearance, we reweight the $\nu_\mu \to \nu_e$ transmutation events by the flux ratio $\Phi_{\nu_e}(E_\nu)/\Phi_{\nu_\mu}(E_\nu)$, while for $\nu_\mu$ disappearance, we make use of a set of $\nu_\mu$ Monte-Carlo events~\cite{MiniBooNE:2022emn}.
We first bin the events by true energy and then degrade the decayed events to lower energy bins, reweighting them with degraded efficiency and cross-section. 
Finally, we perform a similar migration of the true energy distribution to the reconstructed energy distribution as in the appearance channel.
The procedure for antineutrinos is analogous.

In the left panel of~\cref{fig:spectra}, we demonstrate the results of this procedure for a particular parameter point in Model-I and the resulting event rate predicted for MiniBooNE. We choose the parameters $g_\phi = 1.0$, $m_4 = 1$~keV, $|U_{e4}|^2 = 0.10$, and $|U_{\mu 4}|^2=3\times10^{-3}$ to highlight the contribution from sterile neutrino decay in this appearance channel (highlighted in blue).
The oscillation piece of the signal is shown in green. 
In salmon, we show the expected $\nu_e$ disappearance contribution to this event rate (the disappearance/decay of the intrinsic $\nu_e$ background), compared against the expected background in the presence of no new physics shown as a dotted histogram. The upturn of new $\nu_e$ events, especially at low energies compared to the oscillation-only sterile scenario, demonstrates how Model I improves the fit to the MiniBooNE low-energy excess relative to the scenario without decays.

Given a signal prediction in MiniBooNE, we can unfold the event rate excess into an excess of $\nu_e$ in the beam.
Our methodology follows the approach of the first MicroBooNE search for an excess of electron-like events compatible with the MiniBooNE excess~\cite{MicroBooNE:2021tya} and is discussed in greater detail in Ref.~\cite{Arguelles:2021meu}.
Our own implementation of the unfolding procedure is able to reproduce the results of the MicroBooNE collaboration with good accuracy.
In this work, we generalize the unfolding procedure in order to test the most recent MiniBooNE excess reported in \cite{MiniBooNE:2020pnu} (see also \cite{Kamp:2023mjn}).
The intrinsic $\nu_\mu$ and $\nu_e$ backgrounds are modified to account for oscillation/decay events using binned, expected truth-level event rates, then migrated to their expected reconstructed-energy event rates using the MicroBooNE data release associated with~\cite{MicroBooNE:2021nxr}.

Similar to MiniBooNE, we show the predicted signal rate of fully-contained $\nu_e$ events at MicroBooNE (in comparison with the analysis in Ref.~\cite{MicroBooNE:2021nxr}) in the right panel of~\cref{fig:spectra} for the same new-physics parameters adopted in the left panel. 
Again, the relative contributions from $\nu_\mu$ appearance are divided into the oscillation piece (green) and the decay one (blue). 
The dotted histogram shows the expected event rate under the null hypothesis, and the salmon histogram (below the dotted one) demonstrates the disappearance of the intrinsic $\nu_e$ background under this model hypothesis.
The partially-contained events and $\nu_\mu$ sidebands are also modified in an analogous way (except the neutral-current $\pi^0$ sample, which is not oscillated), and are included in the fit.

\section{Discussion}
\label{sec:discussion}

\paragraph{MicroBooNE vs. MiniBooNE} First, we discuss the results for Model (I).
We start with \cref{fig:modelIfit_fixedUe4}, where steriles decay purely via mixing.
The two panels of \cref{fig:modelIfit_fixedUe4} correspond to two separate values for the electron mixing parameter $|U_{e4}|^2$: the first corresponding to approximately the $90\%$ CL upper limit from solar neutrino disappearance ($|U_{e4}|^2 = 0.05$)~\cite{Esteban:2020cvm,Goldhagen:2021kxe} and the second to approximately the best-fit value of the BEST experiment ($|U_{e4}|^2 = 0.1$)~\cite{Barinov:2022wfh}.
We also fix $g_\varphi = 2.5$.
In general, we find that MiniBooNE strongly prefers nonzero mixing over the null hypothesis, whereas MicroBooNE prefers the null hypothesis -- both qualitatively similar to the results of the ``stable'' 3+1 sterile neutrino analyses. Their preferred region (MiniBooNE) and exclusion contours (MicroBooNE) are separated into regions of parameter space dominated by oscillation effects only ($\Delta m_{41}^2 \lesssim 1$ eV$^2$) and those dominated by the decays of $\nu_4$ (larger $\Delta m_{41}^2$), where the boundary depends on the chosen value of $g_\phi$.
The shape of these contours is easy to understand, as in the oscillation-only regime, the $\nu_e$ appearance rate is proportional $|U_{e4}|^2|U_{\mu 4}|^2$ while in the decay regime, it is approximately only proportional to $|U_{\mu 4}|^2$.

The values of $g_\varphi$ and $\Delta m^2_{41}$ help determine the $\nu_4$ lifetime, but once the lifetime is sufficiently smaller than the baseline of the experiment, our fits are independent of their specific values.
This explains the flattening of the MiniBooNE regions at large mass splitting.
The choice of $|U_{e4}|^2$ is still significant: when $|U_{e4}|^2 \gg |U_{\mu 4}|^2$, almost every sterile neutrino produced will decay to a slightly lower energy neutrino state that has large overlap the $\nu_e$ flavor, leading to an effective $\nu_\mu \to \nu_e$ appearance effect, but also $\nu_e$ regeneration for those intrinsic $\nu_e$ in the beam that have disappeared into $\nu_4$.

We repeat a similar exercise for Model II in~\cref{fig:modelIIfit_fixedUe4}, fixing $g_e = 10^{-3}$ and $|U_{e4}|^2 = 0$. Again, we find that the parameter space disfavored by MicroBooNE at $3\sigma$ CL significantly disfavors the preferred parameter space by MiniBooNE. The interplay between the low- and high-$\Delta m_{41}^2$ regions in this parameter space is similar to that of Model I with the caveat that, since $|U_{e4}|^2$ is set to zero, there is no predicted $\nu_\mu \to \nu_e$ or $\nu_e \to \nu_e$ oscillation signature. In the low-$\Delta m_{41}^2$ region, the constraints from MicroBooNE are driven solely by the predicted disappearance of $\nu_\mu$ events. At large $\Delta m_{41}^2$, there is an interplay between $\nu_\mu$ disappearance and $\nu_\mu \to \nu_e$ appearance via sterile decay. 

For both models, generically, we find that the parameter regions that MiniBooNE prefers are significantly disfavored by the current MicroBooNE data. Projecting to the inclusion of future data is challenging -- as evident in both panels of~\cref{fig:modelIfit_fixedUe4}, the constraints derived for MicroBooNE significantly exceed those of the Asimov sensitivity expectation, likely due to the downward fluctuation of the data relative to the expectation. Nevertheless, we summarize the results of these two models, as well as the oscillations-only model (see~\cref{app:3plus1} for further discussion) in~\cref{tab:bestfits}. We provide the best-fit points in each model scenario to our MiniBooNE analysis, as well as the improvement of the MiniBooNE $\chi^2$ at this best-fit point relative to the null hypothesis. Finally, we show the strength with which MicroBooNE \textit{disfavors} each of these best-fit points relative to the expectation of the null hypothesis. 
We see that while these scenarios improve the MiniBooNE test statistic by $\mathcal{O}(30)$ units, MicroBooNE disfavors these with similar strength, $\mathcal{O}(20)$. 
By na{\"i}ve comparison, we can also see that a combined analysis of MiniBooNE and MicroBooNE to these decaying sterile neutrino scenarios would be dominated by MiniBooNE as in the $3+1$ scenario analyzed in Ref.~\cite{MiniBooNE:2022emn}.

Throughout~\cref{fig:modelIfit_fixedUe4,fig:modelIIfit_fixedUe4}, we have fixed the two unseen parameters in demonstrating the measurements as a function of the other two -- we provide more slices of the four-dimensional parameter space of Model I in~\cref{app:MoreSlices}.

\paragraph{MiniBooNE and MicroBooNE compatibility}
In order to measure the compatibility of MiniBooNE and MicroBooNE, we adopt the parameter goodness-of-fit (PGOF) test~\cite{Maltoni_2003}. 
It is obtained by taking the difference between the $\chi^2$ value at the global best-fit from both datasets and the minimum $\chi^2$ values from MiniBooNE and MicroBooNE, 
\begin{equation}\label{eq:PGOF}
    \chi^2_{\rm PGOF} \equiv (\chi^2_{\rm MB}+\chi^2_{\rm \mu B})\big\vert_{\rm min} - (\chi^2_{\rm MB})\big\vert_{\rm min} - (\chi^2_{\rm \mu B})\big\vert_{\rm min}.
\end{equation}
Allowing $g_\phi$ to be non-zero, we observe an improvement in the PGOF from $\chi^2_{\rm PGOF} = 17.7$ to $\chi^2_{\rm PGOF} = 17.4$, although the best-fit still corresponds to an oscillation-dominated scenario. 
This corresponds to $p$-values of $4.9 \times 10^{-4}$ and $1.6 \times 10^{-3}$, under the assumption of Wilks' theorem. This marginal improvement in compatibility is largely due to the additional degree of freedom introduced by the decay model. 
When forcing the solution to be a decay-dominated one, $|U_{\mu 4}|^2 < 0.02$, we find $\chi^2_{\rm PGOF} = 17.59$, which corresponds to a $p$-value of $1.5 \times 10^{-3}$ under Wilks' theorem, suggesting that decays do not improve the compatibility.

\paragraph{IceCube}
The IceCube collaboration has also searched for decaying sterile neutrinos via $\nu_\mu$+$\overline{\nu}_\mu$ disappearance of atmospheric neutrinos~\cite{IceCubeCollaboration:2022tso,weigel-2024}.
While a mild preference for sterile neutrinos with large couplings was found in~\cite{IceCubeCollaboration:2022tso}, the best-fit in the latest analysis~\cite{weigel-2024} is, in fact $g_\phi = 0$, $\sin^22\theta_{24} \simeq 4|U_{\mu 4}|^2 = 0.16$, and $\Delta m^2_{41} = 3.5$~eV$^2$.
While the search focused on decays to invisible particles, thereby neglecting the regeneration of neutrinos after decay, their results are still relevant for the visible-decay models considered here.
This is due to two reasons: i) $\nu_4 \to \hat{\nu}_{s} \varphi$ decays degrade the energy of the daughter neutrino, making them harder to see under the steeply rising atmospheric neutrino background at low energies, and ii) because in model-II (and in model-I when $|U_{e4}|^2 > |U_{\mu 4}|$) the daughter neutrino is predominantly of the electron type, effectively disappearing from the $\nu_\mu$+$\overline{\nu}_\mu$ sample targeted by IceCube.
Therefore, under the justified assumption of negligible $\nu_\mu$ regeneration at IceCube, we can compare our fits and constraints with those of IceCube as illustrated in \cref{fig:modelIfit_fixedUe4} and \cref{fig:modelIIfit_fixedUe4}. We also present a comparison between MiniBooNE, MicroBooNE and IceCube 99\% CL limits with various couplings in \cref{fig:IceCube_comparison}.
It demonstrates that IceCube is able to effectively exclude all of the early-oscillation region of the model, where the oscillation probabilities are proportional to $(L/E)^2$.
However, it is not sensitive to $\Delta m^2_{41}$ values where the visible decay effect is the dominant source of $\nu_\mu \to \nu_e$ appearance events at MiniBooNE. We see overall that IceCube and MicroBooNE are highly complementary in their ability to test this decaying-sterile solution to the MiniBooNE low-energy excess.

\paragraph{MINOS/MINOS+} 
Another constraint on $\nu_\mu$ disappearance is set by MINOS/MINOS+~\cite{MINOS:2017cae}.
It excludes mixing angles as small as $|U_{\mu 4}|^2 \lesssim 1.5 \times 10^{-2}$ at $90\%$ CL in the large $\Delta m^2_{41}$ regime, which corresponds to a neutrino disappearance probability of $1 - P_{\mu \mu} \to 1 - (1 - |U_{\mu 4}|^2)^2 \simeq 0.03$ after all oscillation and decay phase differences have fully developed.
This exclusion seems surprisingly strong compared to the flux and cross-section uncertainties of typical neutrino experiments, which range between $8\%$ and $15\%$~\cite{Louis:2018yeg,Diaz:2019fwt}.
For that reason, we show them as dashed gray lines in our parameter space.
To obtain the limits in the decay models, we follow the data release of Ref.~\cite{MINOS:2017cae} and modify the oscillation probabilities according to \cref{eq:Posc}, neglecting $\nu_\mu$ regeneration effects, which are subdominant for the disappearance channel.
For simplicity, when deriving the limits we allow $\theta_{23}$ to vary but fix $\Delta m^2_{23} = 2.50\times 10^{-3}$~eV$^2$, $\Delta m^2_{21} = 7.54\times 10^{-3}$~eV$^2$, $\sin^2\theta_{12} = 0.277$, 
$\sin^2\theta_{13} = 0.022$,
$\sin^2\theta_{34} = 0$, and $\sin^2\theta_{14} = 0$.
The resulting $90\%$~CL constraints are shown in \cref{fig:modelIfit_fixedUe4,fig:modelIIfit_fixedUe4}, while the $99\%$~CL is shown in \cref{fig:IceCube_comparison}.

\paragraph{Compatibility with BEST and kink searches in $\beta$ decay}
We can test the compatibility with our MiniBooNE and MicroBooNE fits to the parameter space preferred by the BEST experiment, and find significant tension.
We exemplify this tension in~\cref{fig:Ue4sq_vs_m4} assuming the Model-I scenario.
Here, we show constraints/preferred parameter space as a function of the new sterile mass $m_4$ and the mixing with electron flavor $|U_{e4}|^2$. Constraints from $\beta$ decay, reactor antineutrino experiments, and solar neutrino observations (all of which are independent of $|U_{\mu 4}|^2$) as presented in various colors, and the preferred parameter space by the Gallium anomaly (which is mostly independent of the decay coupling $g_{\phi}$ and $|U_{\mu 4}|^2$) is presented in orange.

Sterile neutrinos produced in beta decay can leave a visible kink in the beta electron/positron.
This effect has been searched for in many isotopes~\cite{Dragoun:2015oja,Abdurashitov:2017kka}.
More recently, KATRIN has searched this effect with Tritium, setting leading limits at about $20$~eV and around $100 -700$~eV~\cite{KATRIN:2020dpx,KATRIN:2022ith,KATRIN:2022spi}.

The preferred parameter space for explaining the MiniBooNE low-energy excess is shown at $1$, $2$, and $3\sigma$ CL in closed blue contours in each panel, where we have chosen two representative values of $|U_{\mu 4}|^2$ -- $2.5 \times 10^{-3}$ (top) and $1.0 \times 10^{-2}$ (bottom).
Finally, we demonstrate the power of MicroBooNE in testing these hypotheses. 
For these two representative $|U_{\mu 4}|^2$ values, we find that MicroBooNE can robustly exclude (at $3\sigma$ significance or greater) the strongest-preferred parameter space according to MiniBooNE, and especially its overlap with solutions to the Gallium anomaly.

\paragraph{Cosmology}
The new scalar modifies the behavior of the sterile neutrino in the early universe.
The secret interaction mechanism is automatically triggered by the interaction in \cref{eq:modelI} and, depending on the size of the couplings and masses, the mixing angles in the early universe could be significantly suppressed. 
However, the same interactions can suppress neutrino free streaming during structure formation, leading to a late-time constraint on the model.
While there have been extensive studies on the impact of the new mediator on sterile neutrinos~\cite{Chu:2018gxk,Yaguna:2007wi,Saviano:2013ktj,Giovannini:2002qw,Bezrukov:2017ike,Archidiacono:2020yey,DiValentino:2021rjj,Corona:2021qxl}, the robust predictions that can be made in the context of these short-baseline anomaly solutions is unclear. 
Given the strong constraint from MicroBooNE presented here, there is no pressing motivation to determine the mechanisms by which cosmological observations may be brought into consistency with these sterile neutrino scenarios.
For this reason, we leave cosmological model-building to future work in the event that terrestrial results change to exhibit a strong preference for such decaying sterile neutrinos.

\section{Conclusion}
\label{sec:conclusions}

We presented a new comprehensive MiniBooNE fit and derived new constraints from MicroBooNE on the (visibly) decaying sterile neutrino explanation of the short-baseline anomalies.
We focused on models that can successfully avoid solar antineutrino searches, considering Model-I, a fully sterile-philic scalar particle, and Model-II, a scalar that couples exclusively to electron and sterile neutrinos via higher-dimensional operators.
In both cases, we find regions where the MiniBooNE excess can be explained with the same or greater goodness-of-fit than the minimal sterile neutrino oscillation scenario.
The main differences arise from the neutrino energy degradation in the decay process, which biases the $\nu_e$ signal rate towards lower energies, and from the smaller correlation between $\nu_\mu \to \nu_e$ appearance and $\nu_e$ and $\nu_\mu$ disappearance.
More importantly, however, we find that there are regions of parameter space where constraints from $\nu_\mu$ disappearance do not exclude the MiniBooNE solution.

Using the MicroBooNE $\nu_e$-template search, we derive new constraints on Model-I and Model-II, showing that under decaying-sterile neutrino interpretations, the MicroBooNE data constrains this hypothesis at more than $99\%$ CL.
This limit is typically comparable to or stronger than the one posed in the oscillation-only hypothesis.
In decaying-sterile neutrino solutions, $\nu_\mu$ and $\nu_e$ disappearance effects can be negligible thanks to the smallness of the mixing angles.
This increases the background constraining power of the sideband samples and results in slightly stronger limits. 

In principle, the $\nu_e$ disappearance claimed by the BEST experiment can be explained in Model-I, albeit not without significant tension with solar $\nu_e$ disappearance data.
At higher mass splitting, there is tension with direct searches for sterile neutrinos in the $\beta$ spectrum of tritium and other isotopes.
The IceCube experiment has also searched for decaying-sterile neutrinos, setting constraints on the $|U_{\mu 4}|^2$ parameter by looking for sterile-driven resonances as well as vacuum oscillations in atmospheric $\nu_\mu + \overline{\nu}_\mu$ disappearance.
These complement existing constraints from the MINOS/MINOS+.
Where decay is the dominant effect in MiniBooNE, searches for $\nu_\mu$ disappearance are not sensitive to the model and the MicroBooNE limits become the strongest laboratory-based constraints on the model in a large region of the parameter space.

Finally, we emphasize that the decaying-sterile neutrino scenario is still subject to significant constraints from cosmology and astrophysical observations such as supernovae.
The short-baseline flavor transition we studied complements these probes, providing a direct search for the decay products of the new sterile neutrino.
Future data from the short-baseline program at Fermilab~\cite{MicroBooNE:2015bmn}, including SBND~\cite{Adams:2013gid} and ICARUS~\cite{Antonello:2013ypa}, as well as from JSNS$^2$ in Japan~\cite{JSNS2:2013jdh}, will help clarify the status of the MiniBooNE low energy excess.
Dedicated $\nu_e$-disappearance experiments also play an important role in scenarios like those of Model-I as they can directly constrain $|U_{e4}|^2$, which may be much larger than $|U_{\mu 4}|^2$ in these models.
Future experiments like IsoDAR~\cite{Alonso:2017fci,Winklehner:2023nje} and reactor experiments like DANSS (upgrade)~\cite{DANSS:2018fnn} and PROSPECT-II~\cite{PROSPECT:2021jey} will provide important insight into the decaying sterile solution and complement direct searches for sterile neutrinos with masses above $10$~eV at the KATRIN experiment~\cite{KATRIN:2022ith,KATRIN:2022spi}.

\begin{acknowledgments}
We thank Austin Schneider and Nicholas Kamp for discussions on the MiniBooNE fitting procedure.
The work of MH is supported by the Neutrino Theory Network Program Grant \#DE-AC02-07CHI11359 and the US DOE Award \#DE-SC0020250. KJK and TZ are supported in part by DOE Grant \#DE-SC0010813.
\end{acknowledgments}

\appendix

\section{Impact of increased energy degradation}
\label{app:HF}

\begin{figure}[t]
    \centering
    \includegraphics[width=0.49\textwidth]
    {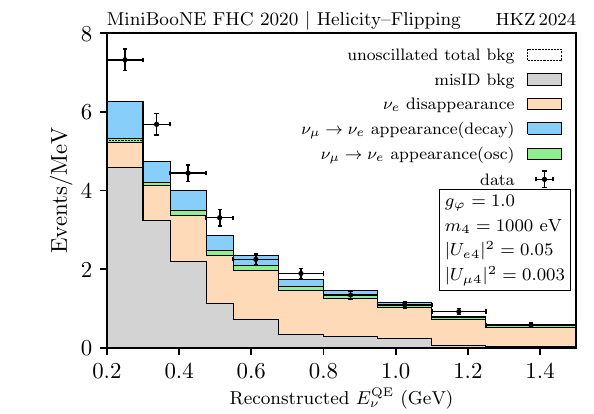}
    \includegraphics[width=0.49\textwidth]
    {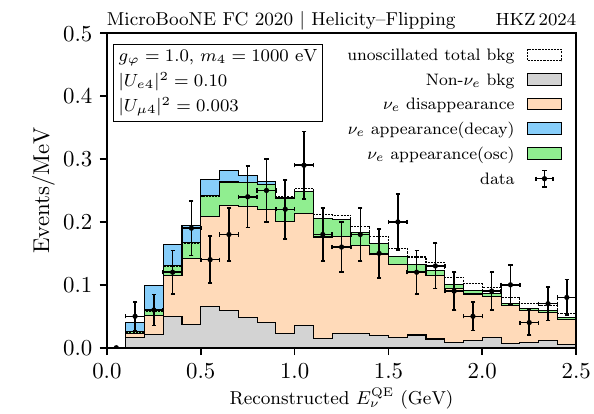}
    \caption{Excess of electron-like events at MiniBooNE(top) and MicroBooNE(bottom) in the helicity-flipping case under the (unphysical) assumption that the daughter neutrinos are detectable. 
    The intrinsic $\nu_e$ contribution to the background is shown as a salmon color and differs from the baseline MiniBooNE expectation (dashed black) due to $\nu_e$ disappearance. The green-filled histograms represent the predicted $\nu_e$ appearance from oscillation, while the blue ones show the contribution from decay (Model-I).}
    \label{fig:event_rate_HF}
\end{figure}

\begin{figure}[h]
    \centering    \includegraphics{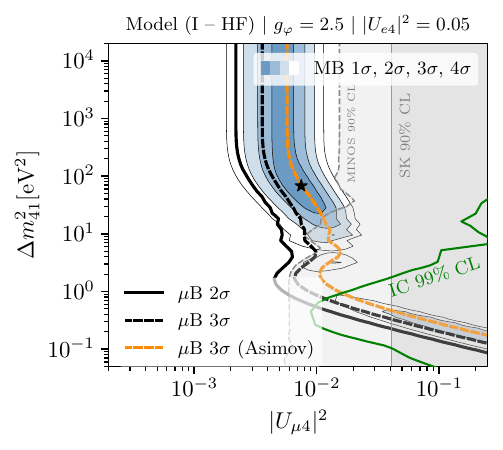}
    \caption{The parameter space of Model (I) for helicity-flipping decays under the (unphysical) assumption that the daughter neutrinos are detectable. 
    We use the same parameters as the left plot in \cref{fig:modelIfit_fixedUe4}. 
    The MicroBooNE constraints are shown as black solid(dashed) lines at $2\sigma$($3\sigma$).
    The green line shows the $99\%$ CL IceCube exclusion region~\cite{weigel-2024}, neglecting $\nu_\mu$ regeneration effects. The grey and light-grey regions show the $90\%$ CL excluded by Super-Kamiokande~\cite{Super-Kamiokande:2014ndf} and MINOS/MINOS+~\cite{MINOS:2017cae}.}
    \label{fig:helicity_flip}
\end{figure}

In this appendix, we comment on the impact of the daughter neutrino energy on MiniBooNE and MicroBooNE fits.
While the decay model we study throughout the main text does degrade the daughter energy somewhat, the daughter neutrino in helicity-conserving (visible) decays involving scalars are emitted preferentially in the direction of the parent particle, leading to a preference for high energy daughters, cf. \cref{eq:dPdX}.
The opposite is true in vector mediator models.
Daughter neutrinos in a helicity-conserving decay (visible) with a vector mediator are emitted preferentially backward with respect to the parent neutrino and so are less boosted in the lab frame.
Neglecting the mass of the mediator in the kinematics, those scenarios lead to a decay distribution that is approximately
\begin{equation}\label{eq:dPdX_2}
    \frac{1}{\Gamma_{\nu_4}} \frac{\dd \Gamma_{\nu_4 \to \nu \phi}}{\dd E_\nu} \simeq \frac{2}{E_4}\left(1 - \frac{E_\nu}{E_4}\right).
\end{equation}
An even stronger preference for lower energy daughter particles can be achieved in scenarios where the boson is close in mass to the parent particle, $m_\phi \lesssim m_4$.

To understand the impact of the decrease in daughter energy on our study, we perform a fit to the MiniBooNE excess and the MicroBooNE data in the scalar model, artificially imposing the decay probability to follow \cref{eq:dPdX_2}.
The resulting energy spectrum in the two experiments can be seen in \cref{fig:event_rate_HF}. 
It prefers noticeably lower energies than in \cref{fig:spectra}.
This would correspond to helicity-flipping decays in Model-I, which leads to invisible right-handed neutrinos in the final state.
While clearly unphysical, performing the MiniBooNE and MicroBooNE fit with this choice helps to quantify the impact of greater energy degradation on our conclusions.
As \cref{fig:helicity_flip} demonstrates, there is an even stronger preference for the decay solution due to the low-energy nature of the MiniBooNE excess, but the MicroBooNE limits remain strong and still exclude the model at greater than $99\%$ CL.
In the plane of \cref{fig:helicity_flip}, we find $\chi^2_{\rm PGOF} = 14.16$, suggesting that greater daughter energy degradation does not substantially improve the compatibility between MiniBooNE and MicroBooNE.

\section{Additional slices of Model-I and Model-II}\label{app:MoreSlices}

\begin{figure*}[h]
    \centering
    \includegraphics[width=\textwidth]{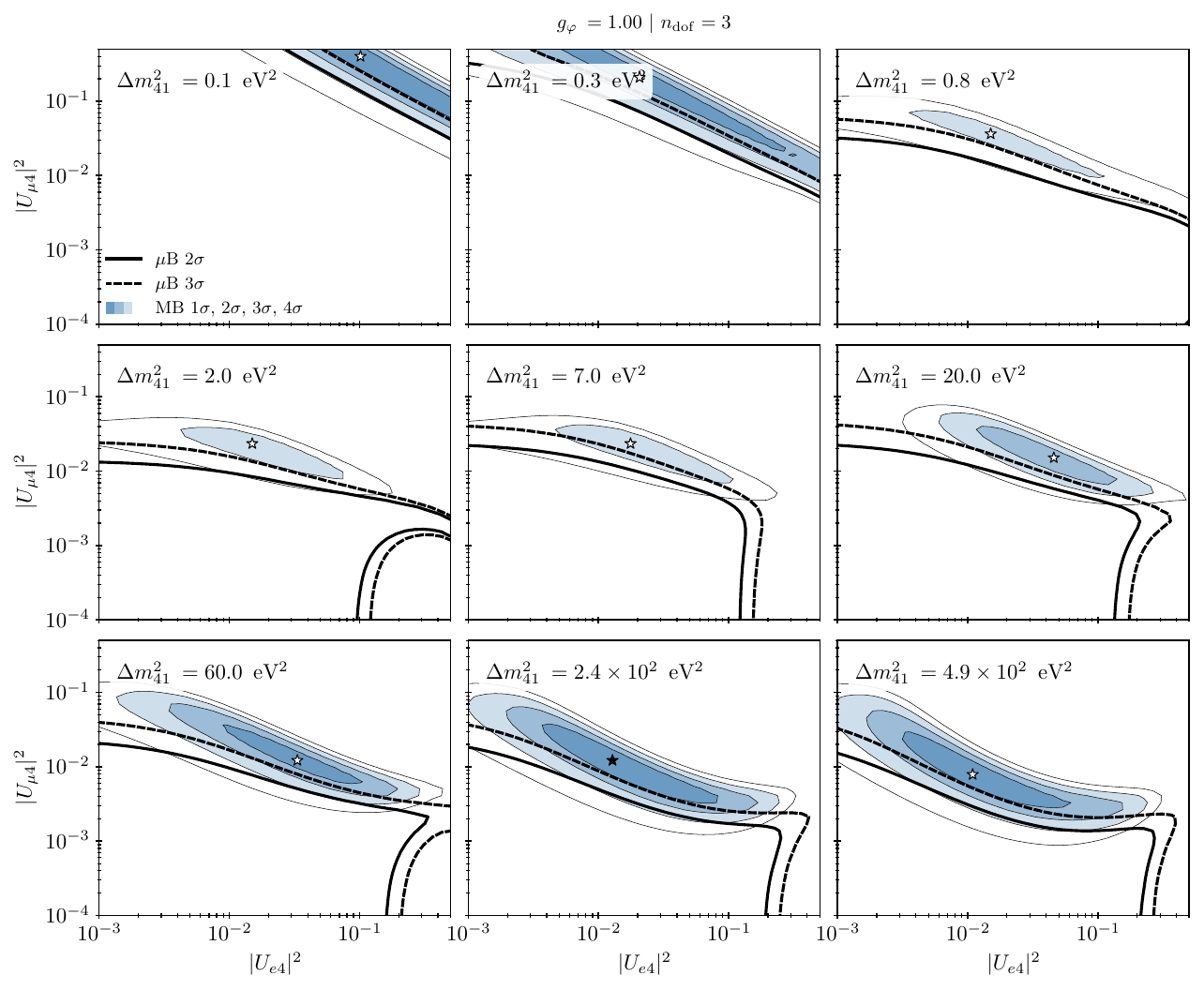}
    \caption{Slices of three-dimensional parameter space of Model-I, fixing $g_\phi = 1$. 
    Each panel corresponds to a fixed $\Delta m^2_{41}$ value and shows the slice in the ($|U_{e 4}|^2$, $|U_{\mu 4}|^2$, $\Delta m^2_{41}$) space. 
    Contours are drawn with respect to the global minimum in the three-dimensional space ($n_{\rm dof} = 3$).
  Open and closed stars correspond to the local and global best-fit of points of MiniBooNE, respectively.
    \label{fig:modelIfit_sliceI}}
\end{figure*}

\begin{figure*}[h]
    \centering
    \includegraphics[width=\textwidth]{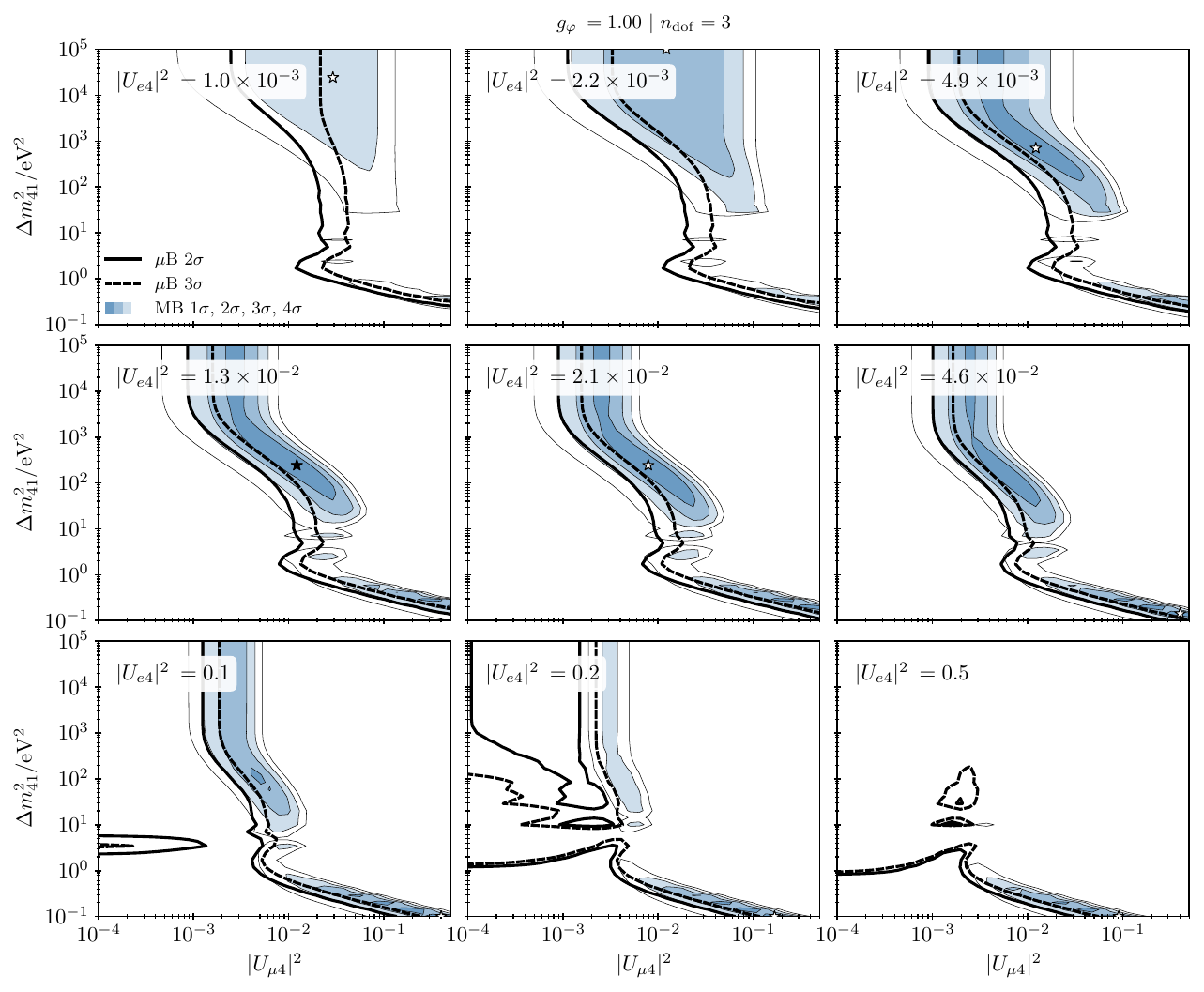}
    \caption{Same as \cref{fig:modelIfit_sliceI} but for slices in the ($|U_{\mu 4}|^2$, $\Delta m^2_{41}$) plane.
    Each panel corresponds to a fixed $|U_{e 4}|^2$ value.
    Contours are drawn with respect to the global minimum in the three-dimensional space ($n_{\rm dof} = 3$).
    Open and closed stars correspond to the local and global best-fit of points of MiniBooNE, respectively.
    \label{fig:modelIfit_sliceII}}
\end{figure*}

This appendix shows additional regions of the decaying sterile neutrino parameter space in Model-I.
\cref{fig:modelIfit_sliceI} shows the results of the MiniBooNE and MicroBooNE fits in slices of $|U_{e4}|^2$ and $|U_{\mu 4}|^2$ of the parameter space for $g_\phi = 1$.
The local best-fit points are shown as open stars and the global best-fit point as a black closed star.
The decay-dominant solution is clearly visible for mass splitting greater than $\Delta m^2_{41} > 20$~eV$^2$.
We also show slices of $|U_{\mu 4}|^2$ and $\Delta m^2_{41}$ for various values of $|U_{e4}|^2$ in~\cref{fig:modelIfit_sliceII}. Larger values of $|U_{e4}|^2$ help reduce the $\nu_4$ lifetime as well as increase the overlap between the daughter neutrino $\hat{\nu}_s$ and $\nu_e$.

\section{Validation and 3+1 oscillations}\label{app:3plus1}

\begin{figure*}[h]
    \centering
    \includegraphics[width=\columnwidth]{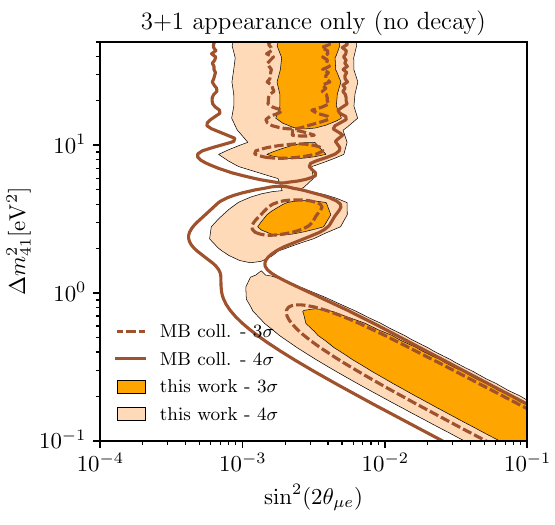}
    \includegraphics[width=\columnwidth]{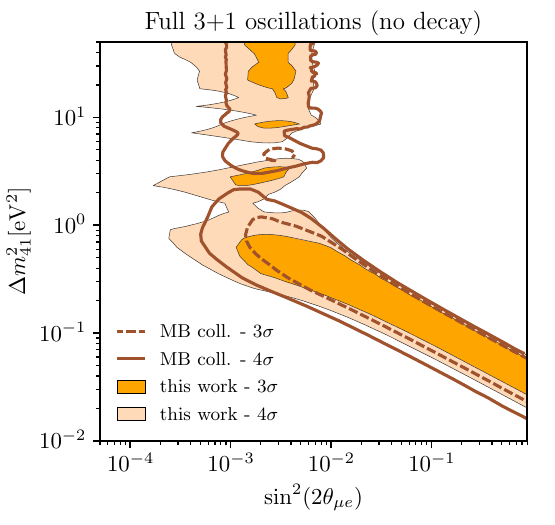}
    \caption{The MiniBooNE preferred regions at $3$ and $4\sigma$ obtained with our fit (orange colors) and by the collaboration (orange lines).
    On the left, we show the fit to $3+1$ oscillations using only appearance, and on the right, we show a fit to full oscillations, including $\nu_e$ and $\nu_\mu$ disappearance, profiling over other parameters.    \label{fig:osc_validation}
    }
\end{figure*}

\begin{figure*}[h]
    \centering
    \includegraphics[width=0.49\textwidth]{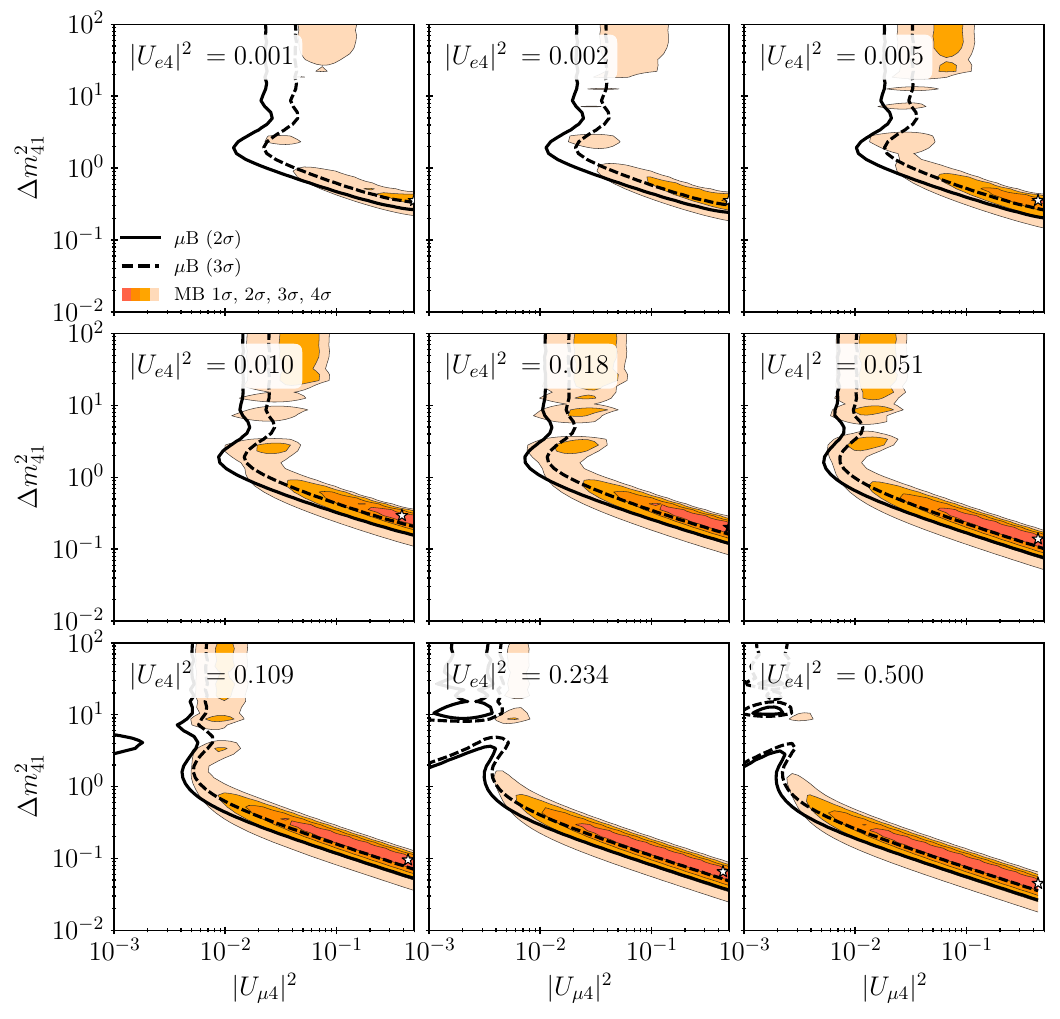}
    \includegraphics[width=0.49\textwidth]{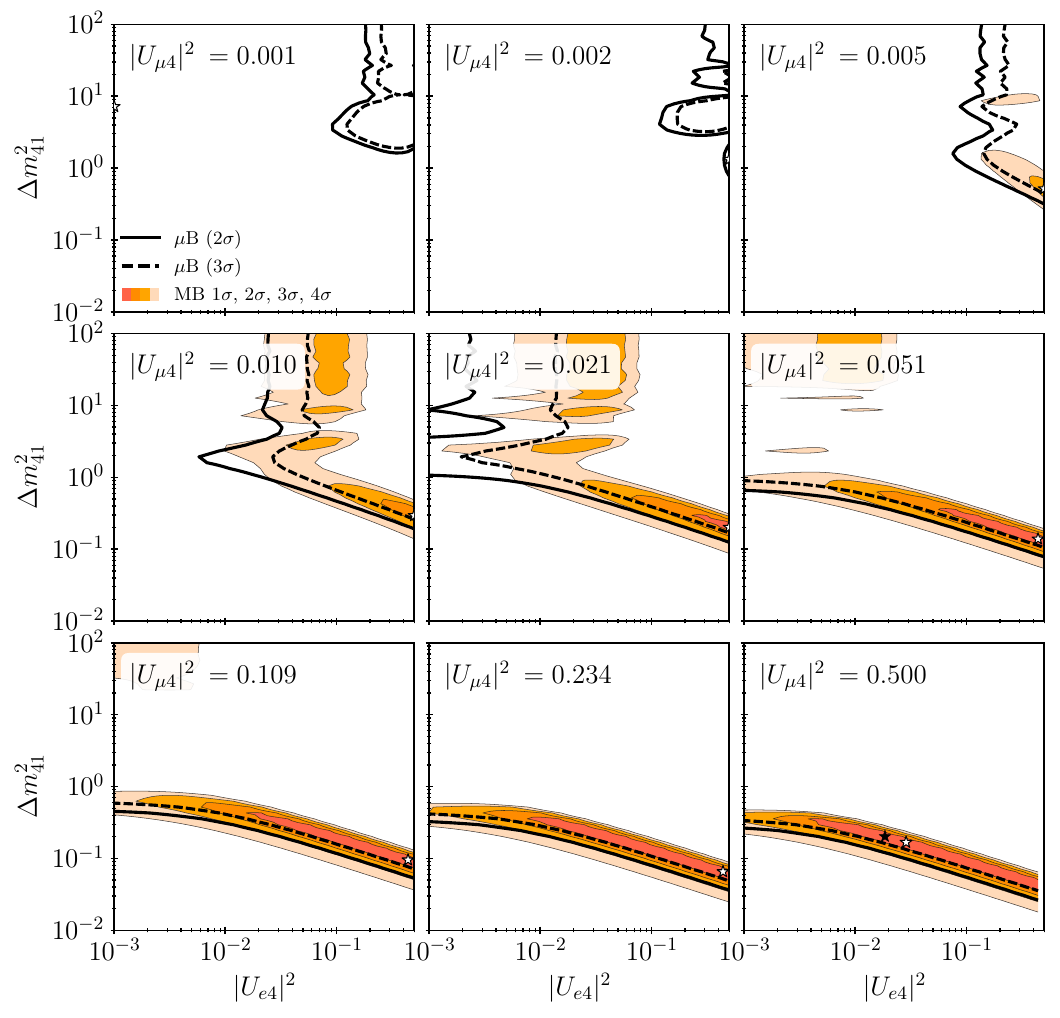}
    \caption{The MiniBooNE preferred regions at $1,2,3 \sigma$ (orange colors) and exclusion limits by MicroBooNE at the $2\sigma$ and $3\sigma$ levels (black lines) in the 3+1 oscillation model without decay.
    Each panel on the left (right) corresponds to a different slice in fixed values of $|U_{\mu 4}|^2$ ($|U_{e 4}|^2$) as indicated within the axes. \label{fig:osc_validation_slices}}
\end{figure*}

In this appendix, we show our fits in a model without sterile neutrino decay, i.e., the standard $3+1$-sterile neutrino model.
We aim to reproduce and further explore the results of the full-oscillation fit performed by the collaboration in~\cite{MiniBooNE:2022emn} using only publicly available data.
A similar study was performed in \cite{Dentler:2019dhz} using only FHC mode data and approximating $E_\nu^{\rm true} \simeq E_\nu^{\rm reco}$ for the $\nu_\mu$ sample, an assumption that is dropped in our analysis thanks to the full $\nu_\mu$ Monte Carlo made available by the collaboration~\cite{MiniBooNE:2022emn}.
In addition, our approach differs from \cite{Dentler:2019dhz} as we do not divide the muon sample by the $\nu_\mu$ disappearance probability.
The justification for doing so in Ref.~\cite{Dentler:2019dhz} was that the prediction for $\nu_\mu$ events was constrained by $\nu_\mu$ data itself.
However, this constraint was applied to the number of pions that decay with a neutrino in the MiniBooNE detector acceptance, increasing it by an overall factor of $f_\pi = 1.22\pm 0.27$~\cite{Aguilar-Arevalo:2008obm}.
Not only does this affect both $\nu_\mu$ and $\nu_e$ samples in similar ways,\footnote{The difference stems from the fact that the $\nu_\mu$ and $\nu_e$ fluxes have similar, but not identical meson parentage.
A larger fraction of intrinsic $\nu_e$ flux comes from $\mu^+$, $K^+$, and $K^0$ decays, in that order of importance, when compared to the $\nu_\mu$ flux.
While the $\mu^+$ component is presumably also subject to the $f_\pi$ normalization, the kaon ones would not be.
This effect is small as the total kaon component is less than half of all the events.
} but it also increases the prediction, as opposed to decreasing it as oscillations would.
This normalization constraint, therefore, cannot be undone by dividing by the oscillation probability.
We opt instead to perform the fit by ignoring this normalization constraint, which in practice means that the constraining power of the $\nu_\mu$ sample is exacerbated in our fit.
We believe this procedure is consistent with the fit performed by the collaboration in Ref.~\cite{MiniBooNE:2022emn}. Our results are demonstrated in~\cref{fig:osc_validation}, which validates the appearance-only scenario (left), where disappearance of intrinsic $\nu_\mu$ and $\nu_e$ is artificially set to zero and in the full 3+1 scenario (right). In both cases we find good, albeit not perfect, agreement with the results of the MiniBooNE collaboration~\cite{MiniBooNE:2022emn}. We then present these results for different slices of $|U_{e4}|^2$ (left) and $|U_{\mu 4}|^2$ (right) in~\cref{fig:osc_validation_slices}.

\bibliographystyle{apsrev4-1}
\bibliography{main}{}

\end{document}